\documentclass[11pt,a4paper]{article}
\pdfoutput=1
\usepackage{jheppub}

\usepackage{amsmath}
\usepackage{verbatim}
\usepackage{amssymb}
\usepackage{inputenc}
\usepackage{textcomp}
\usepackage{appendix}

\newcommand{\e}{\epsilon}
\newcommand{\be}[1]{\begin{equation}\label{#1} }
\newcommand{\ee}{\end{equation}}
\newcommand{\bea}[1]{\begin{eqnarray}\label{#1} }
\newcommand{\eea}{\end{eqnarray}}
\newcommand{\p}{\partial}
\newcommand{\refb}[1]{(\ref{#1})}

\renewcommand{\L}{{\mathcal{L}}}

\newcommand{\Q}{\mathcal{Q}}

\newcommand{\bL}{\bar{{\mathcal{L}}}}
\newcommand{\bQ}{\bar{\mathcal{Q}}}

\renewcommand{\>}{\rangle}

\newcommand{\eps}{\varepsilon}

\renewcommand{\a}{\alpha}
\newcommand{\ta}{\tilde{\alpha}}

\renewcommand{\b}{\beta}
\renewcommand{\t}{\tau}
\newcommand{\s}{\sigma}

\title{Inhomogeneous Tensionless Superstrings}

\author[a]{Arjun Bagchi,} \author[b]{Aritra Banerjee,} \author[c]{Shankhadeep Chakrabortty,} \author[a]{and Pulastya Parekh} \author[]{\\}

\affiliation[a]{Indian Institute of Technology Kanpur, Kanpur 208016, INDIA.\\} 

\affiliation[b]{CAS Key Laboratory of Theoretical Physics, Institute of Theoretical Physics, Chinese Academy of Sciences,  Beijing 100190, CHINA.\\}

\affiliation[c]{Indian Institute of Technology Ropar, Rupnagar, Punjab 140001, INDIA.\\}

\emailAdd{abagchi@iitk.ac.in, aritra@itp.ac.cn, s.chakrabortty@iitrpr.ac.in, pulastya@iitk.ac.in}

\abstract{We construct a novel tensionless limit of Superstring theory that realises the Inhomogeneous Super Galilean Conformal Algebra (SGCA$_I$) as the residual symmetries in the analogue of the conformal gauge, as opposed to previous constructions of the tensionless superstring, where a smaller symmetry algebra called the Homogeneous SGCA emerged as the residual gauge symmetry on the worldsheet. We obtain various features of the new tensionless theory intrinsically as well as from a systematic limit of the corresponding features of the tensile theory. We discuss why it is desirable and also natural to work with this new tensionless limit and the larger algebra.}

\preprint{}

\begin{document}
\maketitle

\section{Introduction}
The massless limit plays an important role in understanding the nature of the theory of particles. In a quantum field theory, the limit explores processes at energies much higher than that of the incoming or outgoing particle states and provides insights into scattering processes at high energies. Massless particles also help understand the causal structure of the spacetime in which they travel. 

Analogously, it is of interest to look at a similar limit in string theory. This is the limit where the tension of the string, $i.e.$ the mass per unit length, becomes zero. The theory of tensionless strings was first explored by Schild in the 1970's \cite{Schild}. Gross and Mende revisited this limit from the perspective of scattering of string states in \cite{Gross:1987kza, Gross:1987ar}. It was found that there were infinitely many relations between string amplitudes of different string states in the tensionless limit \cite{Gross:1988ue} and this is an indication of an emergent very large, hitherto unknown, symmetry as we approach this very high energy sector of string theory. 

There has been recent interest in understanding the tensionless limit of string theory following the discovery of holographic connections \cite{Klebanov:2002ja, Sezgin:2002rt, Gaberdiel:2010pz} between Vasiliev higher spin theories \cite{Vasiliev:2004qz} and free field theories.  It has been speculated that free $\mathcal{N} = 4$ Supersymmetric Yang-Mills theory with a gauge group $SU(N)$ behaves like a theory of tensionless strings \cite{Sundborg:2000wp, Witten-talk}. Other explorations between the connection of higher spin theories with tensionless string theory include \cite{Sagnotti:2003qa, Chang:2012kt, Gaberdiel:2014cha, Gaberdiel:2017oqg}. 

\subsection{Hagedorn Physics and Tensionless Strings} 

Another, albeit less explored, avenue of applicability of the theory of tensionless strings is the physics of the Hagedorn transition. It is well-known that the thermal partition function for free strings diverges above a critical temperature, the Hagedorn temperature $\mathcal{T}_H$, given by 
\be{}
\mathcal{T}_H = \frac{1}{4 \pi \sqrt{\alpha'}}.
\ee
Rather than a limiting temperature beyond which the theory does not exist, this temperature is indicative of a phase transition \cite{Atick:1988si}. Beyond $\mathcal{T}_H$, string theory is thought to transition into the mysterious Hagedorn phase, of which much has been speculated but little firmly established. It has been shown that near the Hagedorn temperature, when one considers the theory of free strings, it is thermodynamically favourable to form a single long string than to heat up a gas of strings \cite{Bowick:1989us,Giddings:1989xe}. When interactions are included, this picture gets more complicated. 

The common folklore regarding treating string theory near the very high energy Hagedorn phase revolves around the breakdown of the description of theory in terms of the string worldsheet. Near the Hagedorn temperature, the strings become effectively tensionless. The effective tension $T_{eff}$ of a string at a temperature $\mathcal{T}$ \cite{Pisarski:1982cn,Olesen:1985ej}: 
\be{}
T_{eff}(\mathcal{T}) = T_0 \ \sqrt{ 1- \left(\frac{\mathcal{T}}{\mathcal{T}_H}\right)^2}
\ee
where $T_0 = (4\pi \alpha')^{-1}$ is the usual string tension. So one would be able to access the physics of the Hagedorn transition by understanding the theory of tensionless strings. Our approach to tensionless string theory has been centred around the emergence of new symmetries in the tensionless limit. One of our principle claims in this programme is that near the Hagedorn temperature, the string worldsheet does not break down, but the symmetries on the worldsheet go from the tensile version to the tensionless version. In the tensionless limit, the worldsheet metric degenerates and we need to make a transition from 2d Riemannian geometry to the language of 2d Carrollian manifolds. We briefly elaborate on this below. So, let us assert again that near the Hagedorn temperature, the worldsheet description of string theory {\em{does not}} break down, it just becomes different. The Hagedorn transition, and the physics associated with it, is extremely important to understand the fundamental structure of string theory. Armed with the new symmetry tools in our analysis, we hope to make substantial headway into this very essential problem. 

\subsection{Tensionless Worldsheet Symmetries} 
Consider tensile free bosonic closed strings. The Polyakov action 
\be{Pol}
S_P = -\frac{T}{2} \int d\sigma d\tau \ \sqrt{-h} \ h^{\alpha \beta} \p_\a X^\mu \p_\b X^\nu g_{\mu \nu} 
\ee
has gauge symmetries for which we need to fix a definite gauge. In the above, $T=\frac{1}{4\pi\a'}$ is the tension of the string, $h_{\a \b}$ is the worldsheet metric and $g_{\mu\nu}$ is the metric on the spacetime in which the string moves. We will take the string to move on flat $D$ dimensional Minkowski spacetime from now on. So, $g_{\mu\nu} = \eta_{\mu\nu}$. The most convenient choice of gauge for the quantisation of relativistic strings is the conformal gauge where 
\be{}
h_{\a \b} = \eta_{\a \b}.
\ee
If the choice of gauge had completely fixed the gauge symmetry, we would have been left with the simple job of quantising free (worldsheet) scalars. However, the conformal gauge does not completely fix the gauge redundancy of the system and we are left with a residual symmetry algebra with which one acts on the Hilbert space to extract the physical spectrum. The residual gauge symmetry turns out to be two copies of the Virasoro algebra, with generators $\L_n$ and $\bL_n$, which satisfies:
\bea{rca}
[\L_n, \L_m] &=& (n-m) \L_{n+m} + \frac{c}{12} \delta_{n+m,0} (n^3-n), \cr
[\L_n, \bL_m] &=& 0, \cr
[\bL_n, \bL_m] &=& (n-m) \bL_{n+m} + \frac{\bar{c}}{12} \delta_{n+m,0} (n^3-n).
\eea
This is, of course, the symmetry algebra of a 2d conformal field theory (CFT). The reason why string theory has been successfully formulated is because the power of the techniques of 2d CFT can be used extensively to quantise relativistic strings. 

The tensionless string action is obtained from the tensile theory by going to the Hamiltonian framework of the theory where in the action, the tension does not appear as a multiplicative constant up front \cite{Isberg:1993av}. In terms of the action \refb{Pol}, the tensionless action is obtained by replacing 
\be{}
T \sqrt{-h} \ h^{\a\b} \rightarrow V^\a V^\b, 
\ee
where $V^\a$ is a vector density. It is clear that this replacement indicates that the metric $h^{\a\b}$ degenerates in this limit. The Riemannian structure of the worldsheet thus undergoes a drastic transformation. The tensionless bosonic string action \cite{Isberg:1993av} becomes 
\be{tless}
S_V = \int d^2 \s \  V^\a V^\b \p_\a X^\mu \p_\b X_\mu . 
\ee
The above action, like the tensile action \refb{Pol}, is invariant under worldsheet diffeomorphisms, and hence there is gauge freedom and a need to fix gauge. In the analogue of the conformal gauge, which in this case becomes 
\be{}
V^\a = (v, 0),
\ee
the action \refb{tless} again has some residual gauge freedom which is given by the algebra 
\bea{gca}
&& [L_n, L_m] = (n-m) L_{n+m} + \frac{c_L}{12} \delta_{n+m,0} (n^3-n), \nonumber\\
&& [L_n, M_m] = (n-m) M_{n+m} + \frac{c_M}{12} \delta_{n+m,0} (n^3-n), \\
&& [M_n, M_m] = 0. \nonumber
\eea
This algebra is the 2d Galilean Conformal Algebra and has found numerous recent applications. This appears as a non-relativistic contraction of the relativistic conformal algebra \refb{rca}, where we systematically take the speed of light $c$ to infinity and hence the name. The GCA has been discussed in the context of the non-relativistic limit of AdS/CFT \cite{Bagchi:2009my}, as emergent symmetries in Galilean gauge theories in the conformal regime \cite{Bagchi:2014ysa,Bagchi:2015qcw}, and also intriguingly as the symmetry algebra governing putative holographic duals of 3d asymptotically flat spacetimes \cite{Bagchi:2010eg} -- \cite{Bagchi:2016geg}. 

 \subsection{Geometry of the Tensionless Worldsheet}
In the context of flat holography, the central idea is the following: flat spacetimes are obtained as infinite radius limits of AdS spacetimes, hence flat holography should also be obtainable from a systematic limit of the well established AdS/CFT correspondence. The infinite radius limit induces a limit on the field theory which is an ultra-relativistic limit ($i.e.$ $c \to 0$) \cite{Bagchi:2012cy}{\footnote{The rather peculiar $c\to 0$ limit has been christened the Carrollian limit after the mathematician Charles Dodgson, who wrote under the pseudonym Lewis Carroll, and his astoundingly imaginative fantasy novel ``Alice's Adventures in Wonderland". }. It turns out that in two spacetime dimensions, the non-relativistic and ultra-relativistic limit yield the same symmetry algebra from the parent 2d CFT. This is because the spacetime now has one contracted and one uncontracted direction \footnote{For more details on this rather confusing statement, please have a look at \cite{Bagchi:2016bcd}.}. 

Singular spacetime limits, like the non-relativistic limit, are characterised by the degeneration of the metric description of the underlying spacetime. Here the spacetime manifold no longer remains Riemannian due to the lack of a non-degenerate metric on the whole of the manifold. The Galilean spacetime manifold is endowed with a Newton-Cartan structure that consists of a contravariant rank 2 tensor $\gamma^{\mu\nu}$ and a time one-form $\tau_\mu$ orthogonal to $\gamma$, i.e. $\gamma^{\mu\nu} \tau_\mu = 0$. In a $D$ dimensional Galilean spacetime, the metric $\gamma^{\mu\nu}$ has $(D-1)$ positive eigenvalues and one zero eigenvalue and denotes the non-dynamical spatial metric at the spatial slices that are orthogonal to the time axis defined by $\t_\mu$. The geometry of the Newton-Cartan manifold is that of a fibre bundle with a 1d base defined by $\t_\mu$ and $(D-1)$ dimensional flat spatial fibres. 

When the speed of light is taken to zero instead of infinity, like in the limit from AdS to flatspace, the spacetime manifold thus obtained is called a Carrollian manifold \cite{Duval:2014uoa, Duval:2014uva, Duval:2014lpa, Hartong:2015usd}. A Carrollian manifold has a structure very similar to a Newton-Cartan manifold, where the base and fibre of the fibre-bundle are interchanged. These structures are dual to each other in the above sense \cite{Duval:2014uoa}. In two dimensions, Newton-Cartan and Carrollian manifolds become identical to each other as the dimension of the base and fibre become the same.  

Let us return to the discussion of the tensionless string. It has been shown that the tensionless limit on the worldsheet is obtained by the following scaling \cite{Bagchi:2013bga}
$$\s \to \s, \ \t \to \eps \t, \ \eps \to 0.$$
This physically represents the string becoming long and floppy as its tension is taken to zero. In terms of the worldsheet, this is thus the limit where the worldsheet speed of light goes to zero. The structure which emerges on the worldsheet is thus that of a 2d Carrollian manifold. The algebra that dictates the theory is given by \refb{gca}.

\subsection{Outlook and outline: Tensionless Superstrings} 
A very natural generalisation of the analysis of \cite{Bagchi:2015nca} is to consider tensionless superstrings. This was done in \cite{Bagchi:2016yyf} and in the following sections, we briefly review our previous construction. A non-exhaustive list of earlier work in this direction is \cite{Gamboa:1989px, Gamboa:1989zc, Lindstrom:1990ar}. So, why another attempt at same problem? 

The symmetry algebra at the core of the construction of \cite{Bagchi:2016yyf} was what we called the Homogeneous Super Galilean Conformal Algebra or SGCA$_H$. This was given by 
\bea{sgcah}
&& [L_n, L_m] = (n-m) L_{n+m} + \frac{c_L}{12} \, (n^3 -n) \delta_{n+m,0}, \nonumber\\
&& [L_n, M_m] = (n-m) M_{n+m} + \frac{c_M}{12} \, (n^3 -n) \delta_{n+m,0}, \\
&& [L_n, Q^\a_r] = \Big(\frac{n}{2} - r\Big) Q^\a_{n+r}, \quad \{Q^\a_r, Q^\b_s \} = \delta^{\a\b} \left[M_{r+s} + \frac{c_M}{6} \Big(r^2 - \frac{1}{4}\Big)  \delta_{r+s,0} \right]. \nonumber
\eea
Here the suppressed commutators between the generators are zero. This is a natural generalisation of the bosonic 2d GCA \refb{gca} and also has interestingly appeared in the study of asymptotic symmetries of flat space supergravity \cite{Barnich:2014cwa}. 

We also mentioned in \cite{Bagchi:2016yyf} that there was another contraction of the two copies of the Super Virasoro algebra which we called the Inhomogeneous Super Galilean Conformal Algebra or SGCA$_I$. The SGCA$_I$, where the fermionic generators were scaled differently (as opposed to the homogeneous scaling of SGCA$_H$), is given by
\bea{sgcai}
&& [L_n, L_m] = (n-m) L_{n+m} + \frac{c_L}{12} (n^3 -n) \delta_{n+m,0}, \nonumber\\
&& [L_n, M_m] = (n-m) M_{n+m} + \frac{c_M}{12} (n^3 -n) \delta_{n+m,0}, \\
&& [L_n, G_r] = \Big(\frac{n}{2} -r\Big) G_{n+r}, \ [L_n, H_r] = \Big(\frac{n}{2} -r\Big) H_{n+r}, \ [M_n, G_r] = \Big(\frac{n}{2} -r\Big) H_{n+r}, \nonumber\\
&& \{ G_r, G_s \} = 2 L_{r+s} + \frac{c_L}{3} \Big(r^2 - \frac{1}{4}\Big)   \delta_{r+s,0}, \ \{ G_r, H_s \} = 2 M_{r+s} + \frac{c_M}{3} \Big(r^2 - \frac{1}{4}\Big)   \delta_{r+s,0}.\nonumber
\eea
In the above, the suppressed commutators are again zero. This algebra was first obtained in the context of supersymmetric extension of the 2d GCA in \cite{Mandal:2010gx}. Clearly, SGCA$_I$ is the richer of the two algebras and an outstanding question is whether a tensionless limit of superstring theory admits this larger algebra as its worldsheet symmetry. In this paper we answer this question in the affirmative. There are other reasons to prefer this as a symmetry algebra over SGCA$_H$, which we discuss in detail later in our paper. 

The following is an outline of the paper. In Sec~\ref{S2}, we revisit the theory of tensile superstrings in order to set up notation and highlight a few aspects that would be useful to us in our tensionless constructions. In Sec~\ref{S3}, we review our earlier work on the homogeneous limit adding a few new ingredients along the way. In Sec~\ref{S4}, we first discuss why it is desirable to work with the inhomogeneous algebra and go on to construct the inhomogeneous tensionless superstring in an intrinsic way starting from an action principle. We recover our answers in Sec~\ref{S3} and Sec~\ref{S4} by means of systematic worldsheet limits from the tensile theory in Sec~\ref{S5}. We end with a section on discussions and future directions. 

\bigskip

\section{Quick look at Tensile Superstrings}\label{S2}

We start with the Ramond-Neveu-Schwarz (RNS) action for a supersymmetric string
\be{taction}
S=-\frac{T}{2}\int d^2\sigma \Big[\p_aX^\mu\p^aX_\mu-i\bar{\psi}^\mu\rho^a\p_a\psi_\mu\Big].
\ee
Here $X^\mu$ are the spacetime coordinates of the string, which are scalars on the worldsheet, and $\psi^\mu$ are their fermionic counterpart, represented by two component Majorana-Weyl spinors denoted by : 
 \be{}
 \psi^\mu=\begin{pmatrix}\psi^\mu_- \\ \psi^\mu_+\end{pmatrix}.
 \ee 
The $\rho^a$ are the worldsheet gamma matrices satisfying the two dimensional Clifford Algebra 
\be{}
\{\rho^a,\rho^b\}=-2\eta^{ab}. 
\ee
An appropriate representation of these matrices is given by :
\be{rhot}
\rho^0=\begin{pmatrix}0 & -i \\ i & 0\end{pmatrix}, \quad\rho^1=\begin{pmatrix}0 & i \\ i & 0\end{pmatrix}.\ee
This corresponds to a Majorana representation where the fermionic fields are real with the fields respecting usual Clifford algebra.
After putting in forms of $\rho$ and switching to light-cone derivatives, the action \refb{taction} takes the form
\be{action_tensile}
S=-\frac{T}{2}\int d^2\sigma \Big[2\p_+X\cdot\p_-X+i(\psi_{+}\cdot\partial_{-}\psi_{+}+\psi_{-}\cdot\partial_{+}\psi_{-})\Big]
\ee
This gives rise to the equations of motion for the fields. Henceforth, unless mentioned otherwise we will be suppressing the Lorentz indices for simplicity.
\be{teom}
\p_-\p_+ X =0, \quad \p_{\pm}\psi_{\mp}=0.
\ee
The action \refb{taction} is invariant under under local diffeomorphism (parameterised by $\xi$) and supersymmetry transformations (parameterised by $\e$). 
Explicitly the following transformations of $X$ and $\psi$ leave the action invariant:
\begin{subequations}\label{susy}
\bea{}
\delta_\xi X&=&\xi^a\p_a X, \quad  \delta_\xi\psi_\pm = \xi^a\p_a \psi_\pm+\frac{1}{2}\p_\pm\xi^\pm\psi_\pm, \\
\delta_\e X&=&\bar{\e}\psi, \quad \delta_\e\psi = -i\rho^a\p_aX\e. 
\eea 
\end{subequations}
The infinitesimal suspersymmetry paramter $\e$ is also a two-component spinor. %, having the structure by convention
 %\be{} \e=\begin{pmatrix}\e_- \\ \e_+\end{pmatrix}. \ee 
The spinor indices $(\pm)$ can be raised and lowered by the charge conjugation matrix ($C=i\rho^0$)\footnote{For details, please have a look at Appendix \ref{ApA}.}. 

The supersymmetry transformations particularly depend on the representation of the Dirac matrices. Using the representation of the $\rho$ matrices \refb{rhot}, the supersymmetry transformations can be written in the following form
\be{tensilesusy}
\delta X =  i \left(\e^{+} \psi_{+}+\e^{-}\psi_{-}  \right) ,\quad \delta \psi_{-} = -2\e^-\partial_{-}X,\quad \delta \psi_{+} = -2\e^+\partial_{+}X. 
\ee
Since these transformations are local, plugging (\ref{tensilesusy}) back into the variation of the action, and requiring $\delta_\xi S=\delta_\e S=0$ separately, gives rise to the following set of  conditions on $\xi^\pm$ and $\e^\pm$:
\be{xiep}
\p_-\xi^+=\p_+\xi^-=0, \quad \p_-\e^+=\p_+\e^-=0.
\ee 

\medskip

\subsection*{Realisation in Superspace}

Now we will see how to realise these transformations (\ref{susy}) in the $\mathcal{N}=(1,1)$ superspace associated to the superstring. We consider Grassmanian coordinates $(\theta,\bar{\theta})$ along with the 2D worldsheet coordinates $\sigma^a$. A general superfield $Y$ can then be constructed as, 
\bea{superspace_t}
Y(\sigma^\pm,\theta,\bar{\theta})%&=&X(\sigma)+\bar{\Theta}\psi(\sigma^\pm)+\bar{\Theta}\Theta B(\sigma^\pm) \\
&=&X(\sigma^\pm)+i\theta\psi_+(\sigma^\pm)+i\bar{\theta}\psi_-(\sigma^\pm)+\frac{1}{2}\bar{\theta}\theta B(\sigma^\pm).
\eea
The auxiliary field $B$ can be put to zero without loss of any generality if we consider on-shell supersymmetry. The total action can the be constructed as a superspace integral 
\be{}
S=-\frac{T}{2}\int d^2\theta d^2\sigma~  \overline{DY} DY.
\ee 
Integrating out the Grassmannian coordinates, one could arrive at the usual RNS action for the superstring sans the auxiliary field term. To see how the coordinate transformations work for the superstring variables, let us start by looking at an arbitrary variation: 
\bea{arbvar}
\delta Y= (\delta\s^+\p_++\delta\s^-\p_-)X &+& i\theta(\delta\s^+\p_++\delta\s^-\p_-)\psi_+ \cr
&&+\delta\theta\psi_+ + i\bar{\theta}(\delta\s^+\p_++\delta\s^-\p_-)\psi_-+\delta\bar{\theta}\psi_-.
\eea
The variation under diffeomorphisms on the superfield $Y$ yields
\begin{subequations}\label{diffsp}
\bea{}
\delta_\xi Y&=&\delta_\xi X+i\theta\delta_\xi\psi_++i\bar{\theta}\delta_\xi\psi_-\\%+\frac{1}{2}\bar{\theta}\theta \delta_\xi B \\
&=&(\xi^+\p_++\xi^-\p_-)X+i\theta(\xi^+\p_++\xi^-\p_-+\frac{1}{2}\p_+\xi^+)\psi_+\nonumber\\
&&\ \ \ \ \ \ \ \ \ \ \ \ \ \ \ \ \ \ \ \ \ \ \ +i\bar{\theta}(\xi^+\p_++\xi^-\p_-+\frac{1}{2}\p_-\xi^-)\psi_-.
\eea
\end{subequations}
Equating \refb{arbvar} to \refb{diffsp} we see that the effect of the diffeomorphism transformations on the superfield $Y$ is equivalent to change of superspace coordinates
\be{}
\delta\s^\pm=\xi^\pm, \quad \delta\theta=\frac{1}{2}\theta\p_+\xi^+, \quad \delta\bar{\theta}=\frac{1}{2}\bar{\theta}\p_-\xi^-.
\ee
Furthermore, imposing the supersymmetry transformations on the superfield $Y$: 
\bea{susysp}
\delta_\e Y&=&\delta_\e X+i\theta\delta_\e\psi_++i\bar{\theta}\delta_\e\psi_-\nonumber\\%+\frac{1}{2}\bar{\theta}\theta\delta_\e B \\
%&=&(-i\e_+\psi_++i\e_-\psi_-)-i\theta(\e_+\p_+X)+i\bar{\theta}(-\e_-\p_-X) \\
&=&(i\e^+\theta)\p_+X+(i\e^-\bar{\theta})\p_-X+i(\e^+)\psi_++i(\e^-)\psi_-. 
\eea
We see again that the effect of supersymmetry transformations on the superfield $Y$ is equivalent to change of superspace coordinates: 
\be{}
\delta\sigma^+=i\e^+\theta, \quad \delta\sigma^-=i\e^-\bar{\theta}, \quad \delta\theta=\e^+, \quad \delta\bar{\theta}=\e^-. 
\ee
Together the transformations (\ref{susy}) realised on the superspace coordinates take the following form:
\begin{subequations}\label{sspace}
\bea{}
\delta\sigma^+&=&\xi^++i\e^+\theta, \ \ \ \delta\theta=\e^++\frac{1}{2}\theta\p_+\xi^+,\\
\delta\sigma^-&=&\xi^-+i\e^-\bar{\theta}, \ \ \ \delta\bar{\theta}=\e^-+\frac{1}{2}\bar{\theta}\p_-\xi^-.
\eea
\end{subequations}

\medskip

\subsection*{Constructing the symmetry algebra}

Now we would like to find the generator of these transformations. However we also need to keep in mind the restrictions on $\xi$ and $\e$ as given by (\ref{xiep}). It translates to $\xi^\pm(\s^\pm)$ and $\e^\pm(\s^\pm)$.  Writing out the variation on the field $Y$ and carefully imposing restrictions on $\xi^\pm$ and $\e^\pm$, we get,
\begin{subequations}\label{sfield}
\bea{}
\delta Y&=&(\delta\sigma^+\p_++\delta\sigma^-\p_-+\delta\theta\p_\theta+\delta\bar{\theta}\p_{\bar{\theta}})Y \\
%&=&\Big[(\xi^++i\e_+\theta)\p_++(\xi^-+i\e_-\bar{\theta})\p_-+\Big(\e^++\frac{1}{2}\theta\p_+\xi^+\Big)\p_\theta+\Big(\e^-+\frac{1}{2}\theta\p_-\xi^-\Big)\p_{\bar{\theta}})Y \\
&=&\Big[\Big(\xi^+(\s^+)\p_++\frac{1}{2}\p_+\xi^+(\s^+)\theta\p_\theta\Big)+\e^+(\s^+)\Big(\p_{\theta}+i\theta\p_+\Big)\nonumber\\
&&\ \ \ +\Big(\xi^-(\s^-)\p_-+\frac{1}{2}\p_-\xi^-(\s^-)\bar{\theta}\p_{\bar{\theta}}\Big)+\e^-(\s^-)\Big(\p_{\bar{\theta}}+i\bar{\theta}\p_-\Big)\Big]Y \\
&=&[\mathcal{L}(\xi^+)+\mathcal{Q}(\e^+)+\bar{\mathcal{L}}(\xi^-)+\bar{\mathcal{Q}}(\e^-)]Y.
\eea
\end{subequations}
Here $\mathcal{L}$'s are the bosonic generators and the  $\mathcal{Q}$'s are the fermionic generators.  Since the diffeomorphism and supersymmetry parameters are functions of worldsheet coordinates, we can write down a Fourier expansion of the form
\be{}
\xi^{\pm} = \sum_{n} a_n^{\pm} e^{in\s^\pm},~~~\e^{\pm} = \sum_{r} b_r^{\pm} e^{ir\s^\pm}.
\ee
We can define the fourier modes of the generators in the following way:
\bea{}
\mathcal{L}(\xi^+)&=&-i\sum_n a^+_n \mathcal{L}_n,~~~\mathcal{Q}(\e^+)=\sum_r b^+_r \mathcal{Q}_r,\nonumber \\
\bar{\mathcal{L}}(\xi^-)&=&-i\sum_n a^-_n \bar{\mathcal{L}}_n,~~~\bar{\mathcal{Q}}(\e^-)=\sum_r b^-_r \bar{\mathcal{Q}}_r,
\eea
which leads to the following form of the generators:
%\begin{subequations}\label{tensgen}
\bea{tensgen}
\mathcal{L}_n&=&ie^{-in\sigma^+}\Big(\p_++\frac{in}{2}\theta\p_\theta\Big),\ \ \ \ \mathcal{Q}_r=e^{-ir\sigma^+}(\p_\theta+i\theta\p_+), \nonumber \\ 
\bar{\mathcal{L}}_n&=&ie^{-in\sigma^-}\Big(\p_-+\frac{in}{2}\bar{\theta}\p_{\bar{\theta}}\Big), \ \ \ \ \bar{\mathcal{Q}}_r=e^{-ir\sigma^-}(\p_{\bar{\theta}}+i\bar{\theta}\p_-). 
\eea
%\end{subequations}
These are the generators of the residual symmetry for the tensile superstring. The expressions above can be used to determine that the symmetry algebra is nothing but two copies of the Super-Virasoro algebra:
\be{}
[\L_n, \L_m] = (n-m) \L_{n+m}, \quad [\L_n, \Q_r] = \left(\frac{n}{2} - r \right) \Q_{n+r}, \quad \{ \Q_r, \Q_s \} = 2 \L_{r+s}. 
\ee 

\bigskip

\subsection*{Mode expansions}

We have seen that varying the action \refb{action_tensile} leads to the equations of motion for the fields \refb{teom}. We now construct the mode expansions in the NS-NS sector that are compatible with \refb{teom}. These are:
\begin{subequations}
\bea{}
X^\mu(\tau,\sigma)&=&x^\mu+2\sqrt{2\a'}\a^\mu_0\tau+i\sqrt{2\a'}\sum_{n\neq 0}\frac{1}{n}\Big[\a^\mu_ne^{-in(\tau+\sigma)}+\ta^\mu_ne^{-in(\tau-\sigma)}\Big], \\
\psi_+^\mu(\sigma,\tau)&=&\sqrt{2\a'}\sum_{r \in \mathbb{Z} + \frac{1}{2}} b^{\mu}_r e^{-ir(\tau+\sigma)}, \quad \psi_-^\mu(\sigma,\tau)=\sqrt{2\a'}\sum_{r \in \mathbb{Z} + \frac{1}{2}} \tilde{b}^{\mu}_{r} e^{-ir(\tau-\sigma)}.
\eea
\end{subequations}
The commutation relations are given by  
\be{modecom}
[X^\mu(\sigma),\dot{X}^\nu(\sigma')]=\eta^{\mu\nu}\delta(\sigma-\sigma'), \quad \{ \psi^\mu_\a(\sigma),\psi^\nu_{\a'}(\sigma') \}=\eta^{\mu\nu}\delta_{\a\a'}\delta(\sigma-\sigma') .
\ee
These lead us to the following brackets:
\be{mode}\begin{split}
[\a_m^\mu,\a_n^\nu]&=[\ta_m^\mu,\ta_n^\nu]=m\delta_{m+n}\eta^{\mu\nu}, \\
\{b_{r}^{\mu},b_{s}^{\nu} \}&=\{\tilde{b}_{r}^{\mu},\tilde{b}_{s}^{\nu} \}=\delta^{\a\a'}\delta_{r+s}\eta^{\mu\nu}. 
\end{split} \ee
We can write down the components of the energy momentum tensor and the super current 
\begin{subequations}
\bea{}
T_{\pm\pm}&=&(\dot{X}\pm X')^2+\frac{i}{2}\psi_\pm\cdot\p_\pm\psi_\pm=0, \\
%T_{--}&=&(\dot{X}-X')^2+\frac{i}{2}\psi_-\cdot\p_-\psi_-=0 \\
J_\pm&=&\psi_\pm\cdot\p_\pm X =0,
\eea
\end{subequations}
which are directly related to the constraints by the relations
\bea{}
\L_n&=&\frac{1}{2\pi\a'}\int_0^{2\pi}d\s\ e^{in\sigma}T_{++},\quad \Q_r=\frac{1}{2\pi\a'}\int_0^{2\pi}d\s\ e^{ir\sigma}J_{+}. 
\eea
%{\bf{INCLUDE:}} Expressions of T and J in terms of $\L, \Q$. 
In terms of modes, the Super-Virasoro constraints are given by
\begin{subequations}
\bea{}
\L_n&=&\frac{1}{2}\sum_m \a_{-m}\cdot\a_{m+n}+\frac{1}{4}\sum_r (2r+n)b_{-r}\cdot b_{r+n}, \\
\Q_r&=&\sum_m \a_{-m}\cdot b_{m+r} .
\eea
\end{subequations}
Using \refb{mode} the Super-Virasoro algebra is generated:
\bea{}
&& [\L_n, \L_m] = (n-m) \L_{n+m} + \frac{c}{12} \delta_{n+m,0} (n^3-n), \crcr
&& [\L_n, \Q_r] = \left(\frac{n}{2} - r \right) \Q_{n+r}, \quad \{ \Q_r, \Q_s \} = 2 \L_{r+s} + \frac{c}{3} \left( r^2 - \frac{1}{4} \right) \delta_{r+s,0}. 
\eea
Here the central terms will arise from the proper normal ordering of the modes, when we quantise the theory. 

\bigskip

\section{Homogenous Tensionless Superstrings} \label{S3}

In this section we would discuss the residual symmetries of the tensionless superstring. We can now write down the supersymmetrised action for a  fundamental tensionless string:  
\bea{tlessaction}
S &=&\int d^2\sigma \left[V^a V^b \p_a X\cdot \p_b X+i\bar{\psi}\cdot\rho^a\p_a{\psi}\right]. 
\eea
This will be the starting point of the supersymmetric version of our fundamental tensionless analysis, inspired by the RNS action \refb{taction}. As mentioned in the introduction, in the tensionless limit, the worldsheet metric $\eta^{ab}$ becomes degenerate and is replaced by a product $V^aV^b$ where $V^a$ is a vector density. A convenient choice of gauge, which is analogous to the conformal gauge in the tensile theory, is $V^a=(1,0)$. The Clifford algebra needs to be redefined in this case:
\bea{modclif}
\{\rho^a,\rho^b\}&=&2V^aV^b.
\eea 
We start with the so-called homogeneous tensionless superstring, which has previously been discussed in some detail in \cite{Bagchi:2016yyf}. A convenient choice of the gamma matices $\rho^a$ to satisfy the modified Clifford Algebra (\ref{modclif}) is simply reducing them to $\rho^a= V^a$. Once $V^a$ is fixed, we have  $\rho^0=\mathbb{I}$ and $\rho^1=\mathbb{O}$. The action (\ref{taction}) is then written as 
\bea{h_action}
S &=&\int d^2\sigma [\dot{X}^2+i\psi\cdot\dot{\psi}] .
\eea
This action is invariant under the following restricted diffeomorphism and supersymmetry:
\begin{subequations}\label{}
\bea{}
\delta_\xi X&=&\xi^a\p_a X, \quad \delta_\xi \psi=\xi^a\p_a \psi +\frac{1}{4}(\p_a\xi^a)\psi, \\
\delta_\e X&=&\bar{\e}\psi, \quad \delta_\e \psi^\mu=-i\rho^a\p_a X \e.
\eea
\end{subequations}
With the chosen $\rho$ matrices, the supersymmetry transformation has the explicit structure as follows
\be{fundahomosusy}
\delta X =i(\e^{-} \psi_{-}+\e^{+}\psi_{+}) , \quad \delta \psi_{-} = -\e^-\dot{X},\quad \delta \psi_{+} = -\e^+\dot{X}.
\ee
Now as done in the case of tensile string, we would require $\delta_\xi S=0$ and $\delta_\e S=0$. This gives us analogous  conditions on $\xi^{0,1}$ and $\e^{\pm}$:
\be{}
\p_0\xi^0=\p_1\xi^1, \quad 
\p_0\xi^1=0, \quad 
\p_0\e^\pm=0.
\ee
The solutions of the above equations gives us the allowed function space for the diffeomorphisms and supersymmetry transformations in general,
\be{homcond}
\xi^0 = f'(\s)\tau+g(\s),~~\xi^1 = f(\s),~~\e^{\pm}=\e^{\pm}(\s).
\ee
We can see how to realise this as a transformation on the superspace by similar methods as we did for the tensile case. We then write down a superfield $Y$ in terms of $X$ and $\psi$, and study the variation in superspace. It leads to the  following superspace translations:
\begin{subequations}\label{homstrans}
\bea{}
\delta\tau&=&\xi^0+i(\e^+\theta+\e^-\bar{\theta})=f'\tau+g+i\e^+\theta+i\e^-\bar{\theta}, \\ \delta\sigma&=&\xi^1=f,\\
\delta\theta&=&\e^++\frac{1}{4}\theta\p_a\xi^a=\e^++\frac{1}{2}f'\theta, \\ \delta\bar{\theta}&=&\e^-+\frac{1}{4}\bar{\theta}\p_a\xi^a=\e^-+\frac{1}{2}f'\bar{\theta}. 
\eea
\end{subequations}
Following the same algorithm as before in tensile case, we write out the transformation of  the superfield Y and impose conditions (\ref{homcond}) on $\xi^a$ and $\e^\pm$ to get
\bea{suphom}
\delta Y&=&(\delta\tau\p_0+\delta\sigma\p_1+\delta\theta\p_\theta+\delta\bar{\theta}\p_{\bar{\theta}})Y. 
\eea
Given the changes in $\tau,\sigma,\theta$ and $\bar{\theta}$ we could expand \refb{suphom} in terms of generators as
\bea{}
\delta Y&=&[L(f)+M(g)+Q(\e^+)+\bar{Q}(\e^-)]Y,
\eea
where $L$ and $M$ are bosonic generators of the algebra and $Q, \bar{Q}$ are corresponding fermionic generators. We can find after similar fourier expansions of the functions $f,~$$g$ and $\e_{\pm}$, that the generators can be written in the following differential form,
\begin{subequations}\label{homogenerators}
\bea{}
L_n&=&ie^{-in\sigma}\left[\p_\sigma+in \Big(\tau\p_\tau+\frac{\theta\p_\theta+\bar{\theta}\p_{\bar{\theta}}}{2}\Big)\right], \quad M_n=ie^{-in\sigma}\p_\tau,\\
Q_r&=&e^{-ir\sigma}(\p_\theta+\frac{i}{2}\theta\p_\tau), \quad \bar{Q}_r = e^{-ir\sigma}(\p_{\bar{\theta}}+\frac{i}{2}\bar{\theta}\p_\tau). 
\eea
\end{subequations}
These generators satisfy the following algebra:  
\bea{SGCAH}
&& [L_m,L_n]=(m-n)L_{m+n}, \qquad [L_m, M_n]=(m-n)M_{m+n}, \nonumber\\
&& [L_m,Q^\a_r]=\left(\frac{m}{2}-r\right)Q^\a_{m+r}, \qquad\{Q^\a_r,Q^{\a'}_s\}=\delta^{\a\a'}M_{r+s}, \nonumber\\
&& [M_m,M_n]=[M_m,Q^\a_r]=0
\eea
which is the classical part of the homogeneous Super Galilean Conformal Algebra or $SGCA_H$ \refb{sgcah}. We note that in the above $Q^\a$ with $\a={1,2}$ denotes $Q, \bar{Q}$. 

\newpage

\section{Inhomogenous Tensionless Superstrings}\label{S4} 
In this section we will move on to the first main result of this paper, that is understanding the theory of inhomogeneous superstrings from the viewpoint of the fundamental tensionless theory. 

\subsection{Why bother?} 
We have already stressed in the introduction that the SGCA$_I$ is the algebra which is the larger and hence richer of the two algebras \refb{sgcah}, \refb{sgcai}. So, it is definitely of interest to try and realise this as symmetries of the tensionless string. In the homogeneous case, the fermions in the theory are somewhat uninteresting. They scale uniformly and hence for the most of the analysis turn out to be essentially spectators. This is particularly true when one attempts to look at the quantum regime of the tensionless superstring. Some of this is evident from our previous analysis in \cite{Bagchi:2016yyf}. The inhomogeneous tensionless superstring is very different in this regard and the fermions play an active role in the tensionless regime. In this paper, we shall mostly be concerned about the classical analysis and would come back to the quantum aspects, like the structure of the vacuum of the tensionless theory, in upcoming work. But this remains one of our principal motivations behind attempting to realise \refb{sgcai} as a symmetry algebra. We shall, however, comment on a rather interesting distinction between the homogeneous and inhomogeneous cases later in the section where we investigate some old claims in the literature about the equivalence of closed and open strings in the tensionless theory building on our bosonic analysis in \cite{Bagchi:2015nca}. 

Apart from the above, and most importantly, as we will elaborate immediately, the inhomogeneous tensionless superstring is the general solution in the space of tensionless superstring theories, while the homogeneous one is a special case. 

\subsection{Representations of the modified Clifford algebra}
As we have repeatedly stressed, the tensionless regime of string theory is characterised by the degeneration of the metric on the worldsheet. This modifies the Clifford algebra \refb{modclif} on the worldsheet. As we discussed above, we need to choose a gauge and a very convenient one is the analogue of the conformal gauge $V^a = (1, 0)$. With this, the modified Clifford algebra (\ref{modclif}) reduces to the matrix equations: 
\be{mcl}
{(\rho^0)}^2 = \mathbb{I},~~{(\rho^1)}^2 = \mathbb{O},~~\rho^0\rho^1+\rho^1\rho^0 = \mathbb{O}.
\ee
One could verify from (\ref{mcl}) that the algebra does not change form under a similarity transformation $\rho^a \to S^{-1}\rho^a S$. The most obvious solution of \refb{mcl} is the trivial solution of $\rho^0 = \mathbb{I}$ and $\rho^1 = \mathbb{O}$, which we have used for the homogeneous tensionless case. Now looking at this trivial solution, one can infer that this particular set of matrices is an isolated conjugacy class of solutions by itself. Taking some general form of matrices and putting them back into the above equation, we can easily see that apart from this trivial solution, there exists a second unique conjugacy class of solutions satisfying \refb{mcl}. A generic form of these matrices have the curious structure
\be{romatrix}
\rho^0= \begin{pmatrix}1 & 0 \\ 0 & -1\end{pmatrix},~~~\rho^1= \begin{pmatrix}0 & a \\ 0 & 0\end{pmatrix}~~\text{or}~~\begin{pmatrix}0 & 0 \\ a & 0\end{pmatrix},
\ee
with $a \in \mathbb{R}$. These matrices are disconnected from the trivial solution by any means of similarity transformations. Doing similarity transformations on \refb{romatrix} one could generate new sets of solutions to \refb{mcl}. However, due to this conjugacy structure the Dirac action and hence, the algebra would not change. We shall choose $a=1$ without any loss of generality and consider the forms:
\be{}
\rho^0=\begin{pmatrix}1 & 0 \\ 0 & -1\end{pmatrix},\quad\rho^1=\begin{pmatrix}0 & 0 \\ 1 & 0\end{pmatrix}.
\ee
This is the choice of gamma matrices on the worldsheet of the tensionless strings that is more general than the trivial choice which led to the homogeneous case. We shall work with this more general representation and analyse what differences this makes for the underlying symmetry structure of the tensionless superstring. 

\subsection{Constructing Inhomogeneous Tensionless Superstrings}

We should note here that in the tensile theory, the spinors are usually Majorana spinors, and hence have real components. But in the case of inhomogeneous limit of the superstring, where fermions are scaled in a non-trivial way, we have no a priori reason to believe that the spinors have to be real. In fact, we will see in a while that the spinors are indeed complex in nature. The spinors $\psi$ are taken to be complex spinors written as $$\psi^\mu=\begin{pmatrix}\psi^\mu_0 \\ \psi^\mu_1 \end{pmatrix}.$$ 
The action (\ref{taction}) written in terms of components now reads:
\be{action_inhom}
S  =\int d^2 \sigma\left[\dot{X}^2+ i(\psi_0^{*}\cdot\dot{\psi_0}+\psi_1^{*}\cdot\dot{\psi_1}-\psi_1^{*}\cdot\psi_0')\right]
\ee
This action has a noticeable asymmetry to it, putting the two complex spinors in completely different footing altogether. \refb{action_inhom} is invariant under the following set of restricted diffeomorphisms:
\begin{subequations}\label{inhomdiff}
\bea{}
\delta_\xi X&=&\xi^a\p_a X \\
\delta_\xi \psi_0&=&\xi^a\p_a \psi_0+\frac{1}{4}\p_a\xi^a \psi_0 \\
\delta_\xi \psi_1&=&\xi^a\p_a \psi_1+\frac{1}{4}\p_a\xi^a \psi_1+\frac{1}{2}(\p_1\xi^0)\psi_0
\eea
\end{subequations}
Notice here the unique form of transformations. This happens because the fermions transform under a combination of diffeomorphism and Weyl transformations as 
\be{}
\delta_\xi \psi=\xi^a\p_a\psi+\frac{1}{4}(\p_a\xi^a)\psi-\frac{1}{4}(\mathcal{E}^{ab}\p_a\xi_b)\rho^0\rho^1\psi,
\ee
where $\mathcal{E}^{ab}$ is the antisymmetric Levi-Civita tensor. It is interesting to note that the last term in the expression plays no role in the tensile transformations as we have (\ref{xiep}) as the conditions on $\xi^\pm$. In the case of the homogeneous limit this term is also absent as $\rho^1=0$. This extra term is crucial for the action to invariant under the transformations and closure of the residual symmetry algebra. Plugging in the explicit forms of $\rho$ matrices, we can write the form of supersymmetry transformations here 
\be{inhomosusyrho}
\delta X =i(\e^{1*} \psi_{0}+\e^{0*}\psi_{1}),\quad 
\delta \psi_{0} = -\e^1\dot{X},\quad
\delta \psi_{1} = -\e^0\dot{X}-\e^1X' .
\ee
As usual we would require $\delta_\xi S=0$ and $\delta_\e S=0$. Considering local transformations, the imposition of invariance gives us the following conditions on $\xi$ and $\e$:
%\bea{}
%\dot{\e_0} =\dot{\e_1^{*}}&=& 0,\nonumber\\
%\e_0 ' = \dot{\e_1} &=& \dot{\e_0^{*}}, \nonumber\\
%\psi_0 &= &\psi_1^{*}
%\eea
\begin{subequations}\label{inhomocondt}
\bea{}
&& \p_0\xi^0=\p_1\xi^1, \, \, \p_0\xi^1=0,\\
&&\p_0\e^{1*}=\p_0\e^0=\p_1\e^1=\p_1\e^{0*},\\
&& \p_0\e^{0*}=\p_0\e^1=0, \, \, \psi_0=\psi_1^{*}.
\eea
\end{subequations}
The last equation gives us a hint about the conjugate structure of the spinor components, about which we would elaborate more from the limiting side of the theory. Indeed, the Majorana representation seems not to be the right one in the case of inhomogeneous limit of the tensionless strings. 

We would like to obtain the generators of the residual symmetry transformations, subject to the above conditions, as we did before in the tensile and the homogeneous tensionless case. Now we have complex spinors and hence the parameters of the supersymmetry transformations $\e$ are also complex. It is thus convenient to work with complex superspace coordinates $(\a,\chi)$. In the next section, from the limiting perspective we will see how they are related to the conventional choice of coordinates $(\theta,\bar{\theta})$. The superfield $Y$ in terms of worldsheet fields $X$ and $\psi$ is now written down as 
\be{ifield}
Y(\sigma^a,\a,\chi)=X(\sigma^a)+i\a\psi_0(\sigma^a)+i\chi\psi_1(\sigma^a).
\ee
We can see how to realise (\ref{inhomdiff}) and (\ref{inhomosusyrho}) as a transformation on this superspace by similar methods as we did before:
\begin{subequations}\label{ihsuspacetr}
\bea{}
\delta\tau&=&\xi^0+i(\e^0\chi+\e^1\a), \\
\delta\sigma&=&\xi^1+i\e^1\chi, \\
\delta\a&=&\e^0+\frac{1}{4}(\p_0\xi^0+\p_1\xi^1)\a+\frac{1}{2}\p_1\xi^0\chi, \\
\delta\chi&=&\e^1+\frac{1}{4}(\p_0\xi^0+\p_1\xi^1)\chi.
\eea
\end{subequations}
 We can use the appropriate expressions of  $\xi^a$ and $\e^a$ by solving (\ref{inhomocondt}): 
\begin{subequations}\label{inhomcond}
\bea{}
\xi^0&=&f(\sigma)'\tau+g(\sigma),\ \ \ \ \xi^1=f(\sigma), \\
\e^0&=&e(\sigma)'\tau+h(\sigma), \ \ \ \ \e^1=e(\sigma), 
\eea
\end{subequations}
where $e$ and $h$ are Grassmann valued functions. The variations of the coordinates under the residual symmetry transformation that preserves the invariance of the action, with the appropriate forms of $\xi$ and $\e$ from (\ref{inhomcond}) are
\begin{subequations}{}
\bea{}
\delta\tau&=&f'\tau+g+ie'\tau\chi+ih\chi+ie\a, \\
\delta\sigma&=&f+ie\chi, \\
\delta\a&=&e'\tau+h+\frac{f'\a}{2}+\frac{f''\tau\chi}{2}+\frac{g'\chi}{2}, \\
\delta\chi&=&e+\frac{f'\chi}{2}. 
\eea
\end{subequations}
Now we can find the variation of the superfield $Y$ due to these transformations and find the corresponding generators 
\be{sfield}
\delta Y=(\delta\tau\p_\tau+\delta\sigma\p_\sigma+\delta\a\p_\a+\delta\chi\p_\chi)Y =[L(f)+M(g)+G(e)+H(h)] Y.
\ee
Collecting the terms, we can see that the generators have the forms,
\begin{subequations}{}
\bea{}
L(f)&=&\Big[f\p_\sigma+f'\Big(\frac{\a\p_\a+\chi\p_\chi}{2}+\tau\p_\tau\Big)+\frac{f''}{2}\tau\chi\p_\a\Big], \\
M(g)&=&\Big[g\p_\tau+\frac{g'}{2}\chi\p_\a\Big], \\
G(e)&=&\Big[e\p_\chi+ie(\a\p_\tau+\chi\p_\sigma)+e'\tau\p_\a+ie'\chi\tau\p_\tau\Big], \\
H(h)&=&\Big[h\p_\a+ih\chi\p_\tau\Big]. 
\eea
\end{subequations}
Since all the parameters $f,g,e$ and $h$ are functions of $\sigma$ we Fourier expand in $e^{in\sigma}$ each of them as we did previously and obtain the Fourier modes of the generators $L_n,M_n,G_r$ and $H_r$. This leads to the following vector fields:
\begin{subequations}\label{inhomgen}
\bea{}
L_n&=&ie^{in\sigma}\Big[\p_\sigma+in\Big(\frac{\a\p_\a+\chi\p_\chi}{2}+\tau\p_\tau\Big)-\frac{n^2}{2}\tau\chi\p_\a\Big], \\
M_n&=&ie^{in\sigma}\Big[\p_\tau+\frac{in}{2}\chi\p_\a\Big], \\
G_r&=&e^{ir\sigma}\Big[\p_\chi+i(\a\p_\tau+\chi\p_\sigma)+ir\tau\p_\a-r\chi\tau\p_\tau\Big], \\
H_r&=&e^{ir\sigma}\Big[\p_\a+i\chi\p_\tau\Big]. 
\eea
\end{subequations}
The above generators satisfy the algebra, which we have been calling the SGCA$_I$:
\bea{agcai} 
&&[L_m,L_n]=(m-n)L_{m+n}, \quad [L_m,M_n]=(m-n)M_{m+n}, \nonumber \\
&&[L_m,G_r]=\Big(\frac{m}{2}-r\Big)G_{m+r}, \quad [M_m,G_r]=\Big(\frac{m}{2}-r\Big)H_{m+r},\ \{G_r,G_s\}=2L_{r+s},\nonumber \\
&&[L_m,H_r]=\Big(\frac{m}{2}-r\Big)H_{m+r}, \quad \{G_r,H_s\}=2M_{r+s}.
\eea
All other suppressed (anti)commutators are zero. Note that this is a classical algebra which will turn into its quantum version \refb{sgcai} when we quantize the theory. 
\subsection{Mode expansions}
Now we discuss the mode expansions of the fundamental inhomogeneous tensionless string. 
%The action for the tensionless limit after we fixed the gauge $V^a=(v,0)$ is 
%\bea{}
%S&=&\int d^2\sigma [\dot{X}^2-\frac{i}{2}(\psi_0\cdot\dot{\psi_1}+\psi_1\cdot\dot{\psi_0}-\psi_0\cdot{\psi'_0})] \\
%S&=&\int d^2\sigma [\dot{X}^2+i(\psi_0^*\cdot\dot{\psi_0}+\psi_1^*\cdot\dot{\psi_1}-\psi_1^*\cdot{\psi'_0})] 
%\eea
%Can we write the spinor as $\psi=\begin{pmatrix}\psi \\ \psi^* \end{pmatrix}$ and now $\psi=\psi_0$ ? Then the action is 
%\bea{}
%S &=&\int d^2\sigma [\dot{X}^2-\frac{i}{2}(\psi^*\cdot\dot{\psi}+\psi\cdot\dot{\psi^*}-\psi\cdot\psi')] ??
%\eea
The equations of motion obtained by varying \refb{action_inhom} wirth respect to $X$ and $\psi$ are:
\be{NSsoln}
\ddot{X}^\mu=0, \quad \dot{\psi}_0^\mu=0, \quad \dot{\psi}_1^\mu={\psi'}_0^\mu. 
\ee
The solutions in the NS-NS sector are of the following form
\begin{subequations}\label{NSsoln}
\bea{}
X^\mu(\tau,\sigma)&=&x^\mu+\sqrt{2c'}B^\mu_0\tau+i\sqrt{2c'}\sum_{n\neq 0}\frac{1}{n}(A^\mu_n-in\tau B^\mu_n)e^{-in\sigma}, \\
\psi_0^\mu(\sigma,\tau)&=&\sqrt{c'}\sum_r \beta^{\mu}_r e^{-ir\sigma}, \\
\psi_1^\mu(\sigma,\tau)&=&\sqrt{c'}\sum_r [\gamma^\mu_r-ir\tau\beta^{\mu}_r] e^{-ir\sigma}. 
\eea
\end{subequations}
In the above, $c'$ is a constant of dimensions of (mass)$^{-1}$ which allows consistency of both sides of the equations.  
Considering (\ref{modecom}) we get the non-zero commutations of the modes:
\be{abb}
[A_m^\mu,B_n^\nu] = 2m\delta_{m+n}\eta^{\mu\nu}, \quad \{\gamma_{r}^{\mu},\beta_{s}^{\nu} \} = 2\delta_{r+s}\eta^{\mu\nu}.
\ee
We note here that the commutation relations of the oscillators are not in a simple harmonic oscillator basis. This would be an important fact when we look at the quantum version of our present construction and attempt to understand the underlying Hilbert space.  We leave that analysis for our upcoming work.

The constraints can be worked out by finding the components of the energy-momentum tensor and the supercurrent from the action, then taking the appropriate derivatives with respect to $\tau$ and $\sigma$. Writing them down explicitly gives us
\begin{subequations}\label{const}
\bea{}
\dot{X}\cdot X'+\frac{i}{4}\Big[\psi_0'\cdot\psi_1+\psi'_0\cdot\psi_1\Big]&=& 4c'\sum_{n} \Big[L_n-in\tau M_n \Big]e^{-in\sigma}=0,\\
\dot{X}^2+\frac{i}{2}\psi_0'\cdot\psi_0&=&4c'\sum_{n}M_ne^{-in\sigma}=0, \\
\psi_0\cdot X'+\psi_1\cdot\dot{X}&=&4c'\sum_{r} \Big[G_r-ir\tau H_r \Big]e^{-ir\sigma}=0, \\
\psi_0\cdot\dot{X}&=&4c'\sum_{r} H_re^{-ir\sigma}=0.
\eea
\end{subequations}
Using these and the mode expansions for the fields, we can have the following expressions for the generators:
\begin{subequations}{}
\bea{}
L_n&=&\frac{1}{2}\sum_{m}  A_{-m}\cdot B_{m+n}+\frac{1}{4}\sum_{r}(2r+n)\Big(\beta_{-r}\cdot\gamma_{r+n}+\gamma_{-r}\cdot\beta_{r+n}\Big),  \\
M_n&=&\frac{1}{2}\sum_{m} B_{-m} \cdot B_{m+n}+\frac{1}{4}\sum_{r}(2r+n)\beta_{-r}\cdot\beta_{r+n},\\
G_r&=&\frac{1}{2}\sum_{m} (A_{-m} \cdot \beta_{m+r} +B_{-m} \cdot \gamma_{m+r}), \\
H_r&=&\frac{1}{2}\sum_{m} (B_{-m} \cdot \beta_{m+r}). 
\eea
\end{subequations}
The classical algebra of the modes of the constraints can be obtained by using the commutation and anti-commutation relations of the oscillators \refb{modecom} which gives us \refb{agcai}. This is the same SGCA$_I$ algebra as we found from the symmetry analysis.

\subsection{Inhomogeneous closed strings as open strings}
One of the very interesting features of our bosonic analysis in \cite{Bagchi:2015nca} was that we could provide evidence to the claim that in the tensionless limit, closed strings behave like open strings \cite{Sagnotti:2011qp}. We focused on the residual symmetry algebra of the tensionless bosonic string, which is the 2d GCA \refb{gca} and used the fact that when we are looking at theories where $c_M=0$, the 2d GCA truncates to a single copy of the Virasoro algebra. This was shown using an analysis of null vectors first in \cite{Bagchi:2009pe}. The residual symmetries for the tensionless closed bosonic string with $c_M=0$ thus reduces to a Virasoro algebra, which is also the symmetries of an open string worldsheet. This is a strong hint that the tensionless closed strings behave as open strings. 

It is natural to expect that such an analysis would be readily generalisable to supersymmetric extensions of the 2d GCA and we would 
be able to make analogous claims for the tensionless closed superstrings. Here we are faced with a rather peculiar problem. 
The homogeneous tensionless superstring with SGCA$_H$ \refb{sgcah} as its symmetry algebra, does not have a Super-Virasoro subalgebra. Naively, it seems a truncation would actually reduce the SGCA$_H$ to a single copy of the Virasoro algebra. This would be very counter-intuitive and would somehow mean that by looking at the tensionless sector of a perfectly fine tensile superstring theory, we somehow are lead to a sector where supersymmetry is no longer preserved. To put things mildly, this would be very peculiar. The SGCA$_I$ \refb{sgcai}, on the other hand, does have a Super-Virasoro sub-algebra. And it is expected that a sector like the bosonic one stated above would lead to a truncation of the algebra down to just a Super-Virasoro algebra. 

We show in Appendix \ref{ApB} that indeed the truncation works in the expected way for the SGCA$_I$. We also find, rather satisfyingly, that through an analysis of null vectors it can be shown that there is no consistent truncation of SGCA$_H$ to a Virasoro algebra. 

So, we see that much like in the case of the bosonic tensionless theory, in the inhomogeneous superstring, the closed strings behave like open strings when we take the tension to zero. This aspect of the homogeneous superstring is somewhat more curious and we hope to come back to this in the near future. But this feature highlights another reason why we are more interested in the inhomogeneous tensionless limit.

\section{Tensionless Limits of the RNS Superstring} \label{S5}
In this section we study the limiting approach where we start from the theory of a closed tensile string and then take an ultra-relativisitc contraction on the worldsheet, i.e. take the worldsheet speed of light effectively to zero. As we have mentioned previously in the introduction, this entails taking particular limits on the worldsheet coordinates, namely 
\be{urscale}\tau\rightarrow\eps \tau~~\text{and}~~\sigma\rightarrow\sigma\ee where $\eps$ is a small dimensionless parameter. We are looking at the tensionless limit and hence $\alpha'$ should also transform as $\alpha'\rightarrow c'/\eps$. %where $c'$ is finite.
 In this method we can make a comparison with the quantities from the fundamental side and relate them with $\eps$. We have to ensure that our action \refb{action_tensile} does not blow up while taking this limit. It turns out that in order to do this, the fermionic degrees of freedom $\psi$'s can be scaled in two different ways: the $\textit{homogenous}$ and $\textit{inhomogeneous}$ scaling. As the name suggests, in the homogeneous scaling we treat both the fermions in the same way while in the latter we scale the fermions differently. The scaling of the fermions also implies that the superspace coordinates $(\theta,\bar{\theta})$ also scales in the same manner inorder to have a meaningful superfield expansion. This specific contraction can be taken at each level, starting from the action, the supersymmetry transformations, the superfield all the way down to the constraint algebra, thus showing how this scaling is consistent with the analysis done in the above sections. 
 
\subsection{Homogeneous Scaling}
Let us mention at first about the homogeneous contraction of the fermions $\psi$:
\be{}
\psi_{\pm,h}=\sqrt{\eps} \ \psi_{\pm,t},
\ee
where the subscript $h$ denotes the homogeneous scaling and the subscript $t$ denotes the tensile theory. This along with the contraction of the worldsheet coordinates \refb{urscale} on the tensile action \refb{action_tensile} gives us the \refb{h_action}, which is the action for the Homogeneous tensionless superstring. To obtain a meaningful contraction of the tensile supersymmetry transformations (\ref{tensilesusy}), we scale the supersymmetry parameters as 
\be{}
\e^{\pm}_{h}=\frac{1}{\sqrt{\eps}}\e^{\pm}_{t}.
\ee
We can also take the limit of the superfield as given in (\ref{superspace_t}). It turns that to preserve the superspace under such a scaling, we have to scale the superspace coordinates in the same way as $\e^\pm$: 
\be{}
\theta_{h}=\frac{1}{\sqrt{\eps}}\theta_{t}~~\text{and}~~\bar{\theta}_{h}=\frac{1}{\sqrt{\eps}}\bar{\theta}_{t}.
\ee
We can apply this scaling all the way from \refb{tensilesusy} to \refb{sspace}, consistently obtaining \refb{fundahomosusy} to \refb{homstrans}\footnote{We also need to consider that the time component of the diffeomorphism parameter scales as $\xi^0\rightarrow \eps\xi^0.$}. We can also apply this scaling to the generators of the residual symmetry transformation of the tensile superstring \refb{tensgen} and obtain the generators for $SGCA_H$ \refb{homogenerators} using the following redefinitions: 
 \bea{}
L_n&=&\L_n-\bL_{-n}, \quad M_n = \eps (\L_n+\bL_{-n}),\nonumber \\
Q_r&=&\sqrt{\eps}\mathcal{Q}_r,\quad \bar{Q}_r = \sqrt{\eps}\bar{\mathcal{Q}}_{-r}.
\eea
It follows once again that these scaled generators follow the SGCA$_H$ algebra. A more detailed exposition of this scaling can be found in \cite{Bagchi:2016yyf, Mandal:2016wrw}.

\subsection{Inhomogeneous Scaling}

As we discussed previously, in the inhomogeneous limit, the components of the spinors would scale differently. We do this in the following way:
\be{iscaling}
 \psi_{0,i}=\eps\psi_{0,t}~~\text{and}~~\psi_{1,i}=\psi_{1,t},\\
\ee
where now the subscript $i$ denotes the inhomogeneous case. Here $$\psi_0=\frac{\psi_++i\psi_-}{\sqrt{2}}, \quad \psi_1=\frac{\psi_+-i\psi_-}{\sqrt{2}}.$$ This ansatz for the scaling is natural given the hints and conditions we picked up through the course of this work. Together with \refb{urscale}, this scaling on the tensile action gives us the form of the action for the Inhomogeneous tensionless superstring:
\bea{}
S&=&\int d^2\sigma [\dot{X}^2+i (\psi_0\cdot\dot{\psi_1}+\psi_1\cdot\dot{\psi_0}-\psi_0\cdot{\psi'_0})] 
%S&=&\int d^2\sigma [\dot{X}^2+i(\psi_0^*\cdot\dot{\psi_0}+\psi_1^*\cdot\dot{\psi_1}-\psi_1^*\cdot{\psi'_0})] 
\eea
This is same as \refb{action_inhom} provided we use the last equation of \refb{inhomocondt}. 
%We can actually work out the limits for \refb{..} to \refb{..} for the inhomogeneous case also. 
Once again we take the proper contraction on (\ref{tensilesusy}) to find out the limiting transformations. Needless to say the structure of the diffeomorphism remain exactly the same. These contractions have the following form
\be{}
\e^0_{i}=\frac{1}{\eps}\e^0_{t},~~~\e^1_{i}=\e^1_{t},
\ee
where $$\e^0=\frac{\e^+-i\e^-}{\sqrt{2}}, \quad \e^1=\frac{\e^++i\e^-}{\sqrt{2}}.$$ The limiting form of the supersymmetry for inhomogeneous contraction has the form
\be{inhomosusylimit}
\delta X =i(\e^0\psi_0+\e^1\psi_1), \quad \delta \psi_{0} = -\e^1\partial_{0}X ,\quad \delta \psi_{1} = -\e^0\partial_{0}X-\e^1\partial_{0}X.
\ee
One notices here that the form of the above supersymmetry transformation is the same as in (\ref{inhomosusyrho}) provided there is a complex conjugate relationship between the components of the supersymmetry parameter, i.e.
\be{}
{\e^{0}}^{*} = \e^1,~~~{\e^{1}}^{*}=\e^0.
\ee
This strengthens our idea of these new conjugate spinors which we have noticed before. It follows that we can perform a similar scaling on the superspace coordinates. We define the complex superspace coordinates in the following way:
\be{} 
\a=\frac{\theta-i\bar{\theta}}{\sqrt{2}},~~~\chi=\frac{\theta+i\bar{\theta}}{\sqrt{2}}.
\ee
Starting from the defination of the superfield in \refb{sfield}, the expression is modified as 
\begin{subequations}
\bea{}
Y&=&X+i\theta\psi_++i\bar{\theta}\psi_- \\
&=&X+\frac{i}{2}(\a+\chi)(\psi_0+\psi_1)+\frac{i}{2}(\a-\chi)(\psi_0-\psi_1) \\
&=&X+i\a\psi_0+i\chi\psi_1
\eea
\end{subequations}
and we end up getting \refb{ifield}. Now this superfield is invariant under the scalings \refb{iscaling} and $\a\rightarrow\eps\a$. We can apply the scaling to the generators $\mathcal{L}_n,\bar{\mathcal{L}}_n$ and $\mathcal{Q}_r,\bar{\mathcal{Q}}_r$ of the worldsheet residual symmetry transformations, as done for the homogeneous case. We can define new generators in the following way
\bea{ingenscal}
L_n&=&\L_n-\bL_{-n}, \quad M_n = \eps (\L_n+\bL_{-n}), \nonumber \\
G_r&=&Q_r-i\bQ_{-r}, \quad H_r = \eps (Q_r+i\bQ_{-r}),
\eea
and obtain the correct expressions of $L_n$, $M_n$, $G_r$ and $H_r$ \refb{inhomgen}, thereby showing the consistency of this proecdure. %One can find the details of the inhomogeneous contraction of the generators in Appendix B. 

\subsection*{Contraction of the mode expansion and the algebra}

Let us now apply the inhomogenios scaling on the mode expansions of the tensile string to see how to obtain the tensionless ones. We need be careful here and should keep in mind that we also have to scale $\a'$ as 
$\a' \rightarrow \frac{c'}{\eps}$. Under these limits, the bosonic modes transform under this scaling as 
\bea{}
X^\mu(\tau,\sigma)&=& x^\mu+2\sqrt{\frac{2c'}{\eps}}\a^\mu_0\eps\tau+i\sqrt{\frac{2c'}{\eps}}\sum_{n\neq 0}\frac{1}{n}\Big[\a^\mu_ne^{-in\sigma}(1-i\eps n\tau)+\ta^\mu_ne^{in\sigma}(1-i\eps n\tau) \Big] \\
&=& x^\mu+\sqrt{2c'}\sqrt{\eps}(\a^\mu_0+\ta^\mu_0)\tau+i\sqrt{2c'}\sum_{n\neq 0}\frac{1}{n}\Big[\frac{1}{\sqrt{\eps}}(\a^\mu_n-\ta^\mu_{-n})-i n\tau\sqrt{\eps}(\a^\mu_n+\ta^\mu_{-n})\Big]e^{-in\sigma}. 
\nonumber
\eea
The fermionic modes scale as
\begin{subequations}\label{}
\bea{}
\psi_0^\mu(\sigma,\tau)&=&\frac{\eps}{\sqrt{2}}\Big(\psi^\mu_++i\psi^\mu_-\Big)=\sqrt{c'}\sum_r \Big[\sqrt{\eps}(b^\mu_r+i\tilde{b}^\mu_{-r})+\eps^{3/2}(\hdots)\Big] e^{-ir\sigma}, \\%(1-i\e n\tau) \\
\psi_1^\mu(\sigma,\tau)&=&\frac{1}{\sqrt{2}}\Big(\psi_+^\mu-i\psi_-^\mu\Big)=\sqrt{c'}\sum_r \Big[\frac{b^{\mu}_r-i\tilde{b}^{\mu}_{-r}}{\sqrt{\eps}}-ir\tau\sqrt{\eps}(b^\mu_r+i\tilde{b}^\mu_{-r})\Big]e^{-ir\sigma}.
\eea
\end{subequations}
If we compare these modes with the one that we obtained intrinsically, it can be easily seen that
\begin{subequations}\label{oscmap}
\bea{}
&& A^\mu_n =\frac{1}{\sqrt{\eps}}\Big(\a^\mu_n-\ta^\mu_{-n}\Big), \quad  B^\mu_n=\sqrt{\eps}(\a^\mu_n+\ta^\mu_{-n}), \\
&& \gamma^{\mu}_r=\frac{1}{\sqrt{\eps}}\Big(b^{\mu}_r-i{\tilde{b}}^{\mu}_{-r}\Big), \quad \beta^{\mu}_r=\sqrt{\eps}(b^{\mu}_r+i{\tilde{b}}^{\mu}_{-r}). \label{fermosc}
\eea
\end{subequations}
 Plugging the above relations back into the constraints give us the connection \refb{ingenscal} between the tensile and the tensionless constraints. The scaling, together with the two copies of the Virasoro Algebra, corresponds to the algebra obtained in \refb{agcai}. Therefore this analysis agrees with the inhomogeneous scaling of the Super-Virasoro algebra to arrive at SGCA$_I$.

\bigskip

\section{Conclusions}

\subsection{A summary of the paper}

In this paper, we have constructed a new tensionless limit of superstring theory. These {\em{inhomogeneous tensionless superstrings}} are characterised by a residual symmetry algebra on the string worldsheet which is larger than the residual symmetries considered in earlier work. This richer algebra is called the Inhomogeneous Super Galilean Conformal Algebra (SGCA$_I$) from which the inhomogeneous tensionless string derives its name. 

After a brief recapitulation to the usual tensile superstrings and the tensionless limit that was previously constructed (which we call the homogeneous tensionless superstrings), we approach the construction of the inhomogeneous tensionless superstrings in two different ways. One in an intrinsic method that starts off with an action which is the ``natural" generalisation of the bosonic tensionless strings to the supersymmetric version. In the tensionless limit, the worldsheet metric degenerates and hence the Clifford algebra for the worldsheet fermions modifies to incorporate this feature. With a choice of representation of this modified Clifford algebra that is different from and more general than the previous (homogeneous) tensionless limit, we find that the analysis of symmetries brings us at a residual symmetry algebra which is larger than the one arrived at in earlier work. We then look at the mode expansions which helps us arrive at the symmetries in a different way. We reproduce our answers of the intrinsic analysis by constructing the inhomogeneous limit by scaling the fermionic fields and corresponding superspace coordinates.  

The emergence of the larger SGCA$_I$ has interesting implications for the quantum aspects of the tensionless theory. In the homogeneous case, the scaling on the worldsheet was such that this practically left the fermions as spectator fields. In the bosonic story, when we looked at the intrinsic analysis of the tensionless case, the mode expansions naturally identified oscillators on the worldsheet which turned out to be Bogoliubov transformations of the usual tensile theory oscillators. This led to some very interesting physics, the most intriguing of which was that there were two distinct vacua, one corresponding to the tensile theory and one to the tensionless theory. These were linked by (singular) Bogoliubov transformations and the tensionless vacuum turned out to be a squeezed state in terms of the tensile vacuum and the tensile oscillators, and hence a very high energy state. We connected this to the appearance of the long string near the Hagedorn temperature and conjectured this to be the worldsheet signature of this long string phase and the onset of the Hagedorn transition. 

For the homogeneous tensionless superstring, the non-participation of the fermions in the tensionless limit meant that the analogous story was not quite true and there were some unnatural differences between the bosonic and fermionic oscillators when mapping from the tensile to the tensionless phases. In the inhomogeneous case, we find a natural extension for our bosonic analysis to the superstring case. We will have more details of this in an upcoming publication. 

In the parameter space of tensionless superstring theories, which are characterised by the representations of the modified Clifford algebra \refb{modclif}, the inhomogeneous tensionless superstring thus corresponds to a generic class of solutions and the homogeneous version is just a special case (corresponding to the trivial solution  $\rho^0 = \mathbb{I}$ and $\rho^1 = \mathbb{O}$). It is thus clear, apart from our motivations for the quantum theory stated above, why we need to study this class of tensionless superstring theories.

\subsection{Comments and future directions}
We have seen that the choice of different gamma matrices in the modified Clifford algebra led us to two different symmetry algebras and two different tensionless superstrings, {\it{viz.}} the homogeneous and inhomogeneous. The natural question is whether this is exhaustive or by other choices of these modified gamma matrices, we could find even more versions. In the gauge $V^a= (1,0)$, these two are the only sectors. We mentioned that the representative element of the matrices that lead to the SGCA$_I$ could change from \refb{romatrix} by similarity transformations, but the algebra remains unaffected. However, in order to prove that these are the only two sectors which emerge in the tensionless limit, one needs to start with a general gauge and not just $V^a= (1,0)$. Recently, spectral flows in these algebras have been studied in \cite{Basu:2017aqn,Fuentealba:2017fck}. It would be of interest to understand whether similarity transformations of these modified gamma matrices can be linked to spectral flows. It would also be of interest to understand whether any transformation or deformation of one of the tensionless theories can lead to the other. At the moment, it seems that these are two disjoint sectors of the tensile theory, albeit one larger than the other. 
%{\bf{[Do spectral flows exist in the homogeneous algebra? Or is this only confined to the inhomogeneous algebra?]}} 

\medskip

There are, of course, five distinct tensile superstring theories and it would be of importance to understand what the tensionless limit means in all these different theories. These theories are also linked by a web of dualities. A study of this web of dualities could be very illuminating in the tensionless limit and may tell us something important in M-theory. 

\medskip

One of the interesting co-incidences of the bosonic tensionless theory was that the symmetry algebra, viz. the bosonic 2d GCA, coincided with the asymptotic symmetries of 3d flat space at null infinity. We observed in \cite{Bagchi:2016yyf}  that even for the homogeneous tensionless superstring, the underlying algebra bore a close resemblance to the symmetries arising out of a canonical asymptotic symmetry analysis of $\mathcal{N}=1$ supergravity in 3d flat space \cite{Barnich:2014cwa} \footnote{The SGCA$_H$ was actually a trivial extension of the asymptotic symmetry algebra of 3d flat supergravity, which was called the super Bondi-Metzner-Sachs algebra in \cite{Barnich:2014cwa}.}. It is of interest to note that the SGCA$_I$ also has recently been realised as the asymptotic symmetries of a particular twisted $\mathcal{N}=2$ flat supergravity \cite{Lodato:2016alv}, which the authors rather wonderfully call ``despotic" supergravity as opposed to the usual ``democratic" limit which leads to the SGCA$_H$  and the usual (untwisted) flat $\mathcal{N}=2$ supergravity. Our previous discussions about the naturalness of the emergence of the SGCA$_I$ in the tensionless limit (with reference to the solutions of the modified Clifford algebra) seems to indicate that in the space of 3d $\mathcal{N}=2$ flat supergravities, the despotic dominates over the democratic theories.

\medskip

A rather fascinating direction in which the theory of tensionless strings has found recent application is the theory of ambitwistors. In \cite{Casali:2016atr}, it was claimed that the null string and the ambitwistor string are one and the same when one considers a particular quantisation of the null string. There also has been some recent advances \cite{Casali:2017zkz} where the connections of this null or ambitwistor string to the Cachazo-He-Yuan (CHY) formalism \cite{Cachazo:2013hca} was explored in further detail.   

\medskip

One of the principal objectives of our project of understanding tensionless strings is to ultimately link up to the results of Gross and Mende \cite{Gross:1987kza, Gross:1987ar} about scattering amplitudes in the very high energy limit of string theory with methods on the worldsheet. With the formalism that we have built in our previous papers and this one, we hope to be able to address this question in the near future. The initial crucial step is defining vertex operators in the 2d GCA and its supersymmetric versions. In the context of the ambitwister string, the authors in \cite{Casali:2017zkz} noted that the theory had $c_M=0$ {\footnote{The authors also show that for the theory they are interested in, $h_M=0$. For more details of what $h_M$ is and on the truncation, see Appendix \ref{ApB}.}}. This meant that the symmetry algebra truncates to a single copy of the Virasoro algebra and one can use usual CFT vertex operators. But for a general theory with a non-zero $c_M$, one would need to find vertex operators of the full GCA and use them for the construction of tensionless string amplitudes.   

\bigskip 

\section*{Acknowledgements}
It is a pleasure to thank Rudranil Basu, Rajesh Gopakumar, Yang Lei, Wout Merbis, Aditya Mehra, V. P. Nair, Marika Taylor and Chi Zhang for helpful discussions. The work of Arjun Bagchi (AB) is supported partially by the Max Planck Society (Germany), the Department of Science and Technology (India). AB thanks the Simon Centre for Geometry and Physics, Centre for Theoretical Physics at MIT, Universite de Libre Brussels (ULB), Vienna University of Technology and the University of Southampton for hospitality at various stages of this work. Aritra Banerjee (ArB) is supported in part by the Chinese Academy of Sciences (CAS) Hundred-Talent Program, by the Key Research Program of Frontier Sciences, CAS, and by Project 11647601 supported by NSFC.  SC acknowledges IIT Kanpur for hospitality during a part of this work. PP thanks IIT Ropar, Erwin Schr{\"o}dinger Institute at Vienna,  ULB and the University of Groningen.

\newpage

\appendix
\section*{Appendices}
\section{Tensionless Spinors} \label{ApA}
In this appendix, we comment on the spinors and conventions related to the spinors used in this paper. 
The first issue to note is about the spinor indices used in this work. An important concept here would be the charge conjugation operator $C$, which acts on the spinor to produce conjugate spinors via the operation
\be{}
\psi \to \psi^c = C\bar\psi^{T}.
\ee
It can be shown that the operator should have the following symmetry by definition,
\be{}
C{(\rho^a)}^{T}C^{-1}=-\rho^a,
\ee
Where $\rho^a$ are the usual Dirac matrices. For example, in case of tensile strings it can be explicitly shown that $i\rho^0$ makes a good choice for a charge conjugation operator.  This has an important consequence as we know that the spinor indices can be raised and lowered via action of the charge conjugation operator in the following way,
\be{}
\chi^A = \chi_B C^{BA},~~~A,B = \pm
\ee
Now we can see that we can raise/lower the indices on tensile spinors as
\be{}
\psi^{+} = \psi_{-},~~~\psi^{-} = -\psi_{+},
\ee
and it could be done similarly for $\e$, the supersymmetry parameter. For the homogeneous contraction, we can again choose the charge conjugation operator as $i\rho^0$ and show that the required identity works out fine. But now the spinor indices are raised/lowered as
\be{lowerupper}
\psi^{+} = i\psi_{+},~~~\psi^{-} = i\psi_{-}.
\ee
This is very useful in seeing how the homogeneous supersymmetry matches with the tensile one under the proper contraction. For the inhomogeneous case, we could also define such a charge conjugation matrix. But, curiously in this case $\rho^0$ does not satisfy the relevant criterion. As
\be{}
\begin{pmatrix}1 & 0 \\ 0 & -1\end{pmatrix}.\begin{pmatrix}0 & 0 \\ 1 & 0\end{pmatrix}.\begin{pmatrix}1 & 0 \\ 0 & -1\end{pmatrix}=\begin{pmatrix}0 & 0 \\ -1 & 0\end{pmatrix},
\ee
we can instead find out another matrix that satisfies the criterion better. Starting out with the set of $\rho$ matrices for the inhomogeneous case, with some tedium, we can propose such a matrix with the form
\be{}
C= \begin{pmatrix}0 & 1 \\ -1 & 0\end{pmatrix}.
\ee
Which has same structure as the charge conjugation matrix for the tensile string. This reaffirms the hints we have picked up throughout the work about the structural similarity of tensile and inhomogeneous Dirac theory.

We can also comment on the chirality of the theory via defining proper projection operators on the spinors. For example, in the case of tensile strings, we can write down the matrix
\be{}
\bar\rho=\rho^0\rho^1=\begin{pmatrix}1 & 0 \\ 0 & -1\end{pmatrix}
\ee
This is analogous to $\gamma_5$ in four dimensions. So the chiral projection operators in two dimensions will have the form $P_{L/R} = \frac{1\pm \bar\rho}{2}$, which picks out the left/right spinor component, but for homogeneous case, the concept completely fails as $\bar\rho = 0$, so that the projection operators are just proportional to unity matrices. In the inhomogeneous case however,
\be{}
\bar\rho=\begin{pmatrix}0 & 0 \\ -1 & 0\end{pmatrix}.
\ee
So also in this case the concept of chiral projection of fermions is not at all clear.

\bigskip \bigskip

%[\textbf{Note: To take the homogeneous limit correctly, one should take the limit on \ref{propersusy} (tensile) to arrive at 
%\bea{tensilelimitsusy1} 
%\delta X &=&  i \left(\e^{+} \psi_{+}+\e^{-}\psi_{-}  \right) ,\nonumber \\
%\delta \psi_{-} &=& \partial_{0}X \e^{-} ,\nonumber\\
%\delta \psi_{+} &=& \partial_{0}X \e^{+} 
%\eea
%Now compare it with \ref{fundahomosusy} (fundamental homogeneous) and change the indices on that equation with $\e$ lowered/raised according to \ref{lowerupper} and they would match exactly. This confusion just happens due to ill defined spinor index structure during the translation from tensile theory and its homogeneous contraction!}]

\section{Truncation of the Algebra}\label{ApB}

In \cite{Bagchi:2009pe}, it was shown how GCA can be reduced to a single copy of Virasoro Algebra, provided one of the the central charges, $c_M$ is zero. This is done by analysing null vectors (states that are orthogonal to all states in the Hilbert space including themselves) of the GCFT. 

The recipe is as follows: We define general states by considering a linear combinations of the descendants ($L_{-n}, M_{-n}$ etc.) acting on the primary state labeled by the weights $h_L$ and $h_M$. Then we impose the conditions that all the positive modes anhilate this state, which fixes the coefficients for the linear combination of the null states along with a relation between the weights and the central charges. Since we consider $c_M=0$, we find that we have non-trivial null states only for $h_M=0$. It is sufficient to perform this analysis upto level 2. We see that the null states at each level consists only the $M$ modes. Now if we remove the null states from the Hilbert Space, we are left with only the $L$ states, which belong to the representation of a single copy of the Virasoro Algebra. 

We are going to attempt to do the same analysis of null states for the two versions of the Supersymmetric GCA that we have in conncetion to tensionless superstrings. This time we need to consider fermionic generators as well as general states with half-integral levels (as we are dealing with states in the NS-NS sector).

\subsection*{Null states in $SGCA_I$ in the NS-NS Sector}
Let us consider representations of the inhomogeneous algebra \refb{sgcai}. We will label the states with the eigenvalues of $L_0$, which we will call $h_L$. Since $[L_0, M_0]=0$, this means that these states are further labeled under $M_0$. So,
\be{}
L_0 |h_L,h_M\rangle = h_L |h_L,h_M\rangle, \quad M_0 |h_L,h_M\rangle = h_M |h_L,h_M\rangle
\ee
Since we are in the NS-NS sector, the supersymmetry generators $G_r, H_s$ have only half integer modes and we don't have any extra labels compared to the bosonic case. We construct the notion of primary states $|\phi_p\> \equiv |h_L, h_M\>_p$ in the theory such that the $h_L$ value is bounded from below. The positive modes annihilate the primary state:
\begin{equation}\label{anh}
\begin{split}
L_n|\phi_p \rangle&=M_n|\phi_p \rangle=G_r|\phi_p \rangle=H_r|\phi_p \rangle=0 \ \ (n,r>0),
\end{split}\ee
while the negative modes raise the $h_L$ eigenvalue. 

Below we make a list of a general state at an arbitrary integral or half-integral level $N$ and the corresponding null state $|N\>$ after solving the conditions we obtain by imposing that all positive modes annihilate them. It is trivial to see that $L_n|N\rangle=0$, provided $n>N$. The same is applicable for $M_n,G_r$ and $H_r$ where $n$ or $r>N$. Thus we only need to calculate upto number of conditions where $n$ or $r=N$. 

\paragraph{Level 1/2:} We can write down the most general state $|1/2\>$ at this level by considering a combination of the fermionic generators only: 
\be{} \begin{split}
&a_1G_{-\frac{1}{2}}|\phi_p \rangle+a_2H_{-\frac{1}{2}}|\phi_p \rangle. \\  
\end{split}\ee
Now we impose the condition that $G_{\frac{1}{2}}$ and $H_{\frac{1}{2}}$ anhilate this state. This will give us some conditions on the coefficients $a_1$, $a_2$ and the weight $h_M$. We can solve for the coefficients to obtain the null state at this level. It is easy to see that we need to simply solve for 
\be{}
a_1h_L+a_2h_M=0,\quad a_1h_M=0.
\ee
To get a non-trivial state we must have $h_M=0$, and therefore the null state is 
\be{} \begin{split}
|1/2\rangle&=a_2H_{-\frac{1}{2}}|\phi_p \rangle. \\  
\end{split}\ee

\paragraph{Level 1:} 
The most general state:
\be{} b_1L_{-1}|\phi_p \rangle+b_2M_{-1}|\phi_p \rangle+b_3G_{-\frac{1}{2}}H_{-\frac{1}{2}}|\phi_p \rangle.\ee
 We impose the anhilation conditions %$G_{\frac{1}{2}}|1\>=H_{\frac{1}{2}}|1\>=L_{1}|1\>=M_{1}|1\>=0$
 and obtain $b_1=b_2h_M=b_3h_M=0,\ b_2+2b_3h_L=0$. To get a non-trivial state we must have $h_M=0$, and the level $1$ null state is 
\be{} \begin{split}
|1\rangle&=b_2\Big[M_{-1}-\frac{1}{2h_L}G_{-\frac{1}{2}}H_{-\frac{1}{2}}\Big]|\phi_p\>. \\  
\end{split}\ee
The $2^{nd}$ term is the descendant of the null state at level $1/2$. So if we set $|1/2\rangle$ along with is descendants to zero (in order to remove level $1/2$ null states from the Hilbert Space), then we are left with $M_{-1}|\phi_p \rangle$ as the null state at level $1$. 

\paragraph{Level 3/2:} 
The most general state:
\be{} \begin{split}
&\ \ d_1L_{-1}G_{-\frac{1}{2}}|\phi_p \rangle+d_2L_{-1}H_{-\frac{1}{2}}|\phi_p \rangle+d_3M_{-1}G_{-\frac{1}{2}}|\phi_p \rangle\\
&+d_4M_{-1}H_{-\frac{1}{2}}|\phi_p \rangle+d_5G_{-\frac{3}{2}}|\phi_p \rangle+d_6H_{-\frac{3}{2}}|\phi_p \rangle. \\  
\end{split}\ee
If we concentrate only on the case where $c_M=0$, we will find that to get a non-trivial state at this level, we need to set $h_M=0$. This would mean $d_1=d_5=0$. Then the level $\frac{3}{2}$ null state would be
\be{} \begin{split}
|3/2\rangle&=\Big[d_2\Big(L_{-1}H_{-\frac{1}{2}}+M_{-1}G_{-\frac{1}{2}}-(1+h_L)H_{-\frac{3}{2}}\Big)+d_4M_{-1}H_{-\frac{1}{2}}\Big]|\phi_p \rangle. \\  
\end{split}\ee
Except for the $3^{rd}$ term, all the other terms are descendants of the null state at level $1/2$ and $1$. As seen before, if we set the null states $|1/2\rangle=|1\rangle=0$ along with their descendants, we are left with $H_{-\frac{3}{2}}|\phi_p \rangle$ in the null state at level $3/2$. 

\paragraph{Level 2:} 
At level $2$, the general state is given by a linear combination of 9 terms 
\be{null2} \begin{split}
&\Big[f_1L_{-2}+f_2L^2_{-1}+f_3L_{-1}M_{-1}+f_4M^2_{-1}+f_5M_{-2}+f_6L_{-1}G_{-\frac{1}{2}}H_{-\frac{1}{2}}\\
&\hspace{1.7cm}+f_7M_{-1}G_{-\frac{1}{2}}H_{-\frac{1}{2}}+f_8G_{-\frac{3}{2}}H_{-\frac{1}{2}}+f_9G_{-\frac{1}{2}}H_{-\frac{3}{2}}\Big]|\phi_p \rangle. \\  
\end{split}\ee
We find a set of equations which we can solve to obtain the non-trivial null state at this level. If we set $c_M=0$, these are:
\be{} \begin{split}
f_1=f_2&=0, \quad f_3h_M=f_8h_M=f_9h_M=0, \\  
f_3+2f_4(1+h_L)+2f_8&=0,\quad f_3-2f_7h_M-2f_9=0, \\
2f_5+f_7(3+2h_L)&=0,\quad \frac{3f_6}{2}+f_9(3+2h_L)=0, \\  
(6f_3+4f_6+6f_7+5f_8+3f_9)h_M&=0,\quad f_3+2(f_4+f_7h_M+f_8)=0, \\
f_3+f_4(1+2h_L)+\frac{5f_6}{4}+2f_7h_M+&f_8\Big(h_L+\frac{c_L}{3}\Big)+2f_9=0, \\
2f_3(1+h_L)+2(f_5+f_7)h_M+3f_6+2f_9&=0,\quad 2f_4(1+h_L)+2f_7h_M+2f_8+2f_9=0.  
\end{split} \nonumber\ee
Demanding a non-trivial solution, we can simplify the above equations by $h_M=f_1=f_2=0$, $f_3=2f_9$. For $h_L\neq0$ and $c_L\neq\frac{9}{2}$,  %$f_8=f_9=-\frac{f_4}{2}\neq0$ (provided $h_L=1$). If $h_L\neq1$ then $h_M=f_1=f_2=f_3=f_4=f_6=f_8=f_9=h_M=0$. 
the null state at this level is:

\be{} \begin{split}
|2\rangle&=f_3L_{-1}M_{-1}|\phi_p \rangle %+f_4L_{-1}G_{-\frac{1}{2}}H_{-\frac{1}{2}}|\phi_p \rangle\\
+f_7\Big[M_{-1}G_{-\frac{1}{2}}H_{-\frac{1}{2}}|\phi_p \rangle-\frac{3+2h_L}{2}M^2_{-1}|\phi_p \rangle\Big] \\
&+f_8\Big[G_{-\frac{3}{2}}H_{-\frac{1}{2}}|\phi_p \rangle-G_{-\frac{1}{2}}H_{-\frac{3}{2}}|\phi_p \rangle+\Big(2+\frac{4h_L}{3}\Big)M_{-2}|\phi_p \rangle\Big] \\
\end{split}\ee
Except for $M_{-2}|\phi_p \rangle$, all the remaining terms are descendants of the null states at lower levels. So, the only new null state at level 2 is 
$M_{-2}|\phi_p \rangle$. We can thus set $M_{-2}|\phi_p \rangle$ and its descendants to zero and carry out the same exercise for arbitrary higher levels. This means we can throw away the $H$ s at all half-integer levels and the $M$ s at integer levels. This essentially truncates the algebra to $L$ and $G$s, leaving us with a single copy of Super Virasoro algebra.

\subsection*{Null states in $SGCA_H$ in the NS-NS Sector}

Let us look at a similar analysis for the other algebra \refb{SGCAH}. We consider a similar primary state $|\phi_p\>=|h_L,h_M\rangle$, where $h_L$ and $h_M$ are the weights of $L_0$ and $M_0$ correspondingly. %($L_0|\phi_p \rangle=h_L|\phi_p \rangle$ and $M_0|\phi_p \rangle=h_M|\phi_p \rangle$).
 We can also write down the descendants by acting $L$, $M$, $Q$ and $\bar{Q}$ on $|\phi_p\>$, and define their weights under $L_0$ 
\begin{equation}\label{rep}
\begin{split}
L_0L_{-n}|\phi_p \rangle&=(h_L+n)L_{-n}|\phi_p \rangle\ (n>0), \\
L_0M_{-n}|\phi_p \rangle&=(h_L+n)M_{-n}|\phi_p \rangle\ (n>0), \\
L_0Q^\a_{-r}|\phi_p \rangle&=(h_L+r)Q^\a_{-r}|\phi_p \rangle\ (r>0). 
\end{split}\ee
The anhilation conditions are
\be{} L_n|\phi_p \rangle=M_n|\phi_p \rangle=Q_r|\phi_p \rangle=\bar{Q}_r|\phi_p \rangle=0 \ \ (n,r>0). \ee
Now we proceed to do the null state analysis as we did for the previous case. We can mention again that it is trivial to see that $L_n$, $M_n,Q_s$ and $\bar{Q}_s$ acting on a state $|m+r\rangle0$ gives zero, provided $n$ or $s>m+r$. Once again, we can list the null states level by level and analyse the corresponding null state $|p\>$ after imposing the anhilation conditions. Just as before, will be required to set $h_M=0$ at all levels in order to get a non-trivial null state.

\paragraph{Level 1/2:} The general state is $a_1Q_{-\frac{1}{2}}|\phi_p \rangle+a_2\bar{Q}_{-\frac{1}{2}}|\phi_p \rangle$. We then impose the conditions that $Q_{\frac{1}{2}}$ and $\bar{Q}_{\frac{1}{2}}$ anhilate this state. This gives us  $a_1h_M=a_2h_M=0$, and therefore the null state is 
\be{} \begin{split}
|1/2\rangle&=\Big[a_1Q_{-\frac{1}{2}}+a_2\bar{Q}_{-\frac{1}{2}}\Big]|\phi_p \rangle. \\  
\end{split}\ee

\paragraph{Level 1:} 
The general state is given by 
\be{} b_1L_{-1}|\phi_p \rangle+b_2M_{-1}|\phi_p \rangle+b_3Q_{-\frac{1}{2}}\bar{Q}_{-\frac{1}{2}}|\phi_p \rangle. \ee The conditions obtained are : $b_1=b_3h_M=b_2h_M=0$, and the level $1$ null state is 
\be{} \begin{split}
|1\rangle&=\Big[b_2M_{-1}+b_3Q_{-\frac{1}{2}}\bar{Q}_{-\frac{1}{2}}\Big]|\phi_p \rangle. \\  
\end{split}\ee

\paragraph{Level 3/2:} 
The general state is given by
\be{} \begin{split}
&\ \ d_1L_{-1}Q_{-\frac{1}{2}}|\phi_p \rangle+d_2L_{-1}\bar{Q}_{-\frac{1}{2}}|\phi_p \rangle+d_3M_{-1}Q_{-\frac{1}{2}}|\phi_p \rangle\\
&+d_4M_{-1}\bar{Q}_{-\frac{1}{2}}|\phi_p \rangle+d_5Q_{-\frac{3}{2}}|\phi_p \rangle+d_6\bar{Q}_{-\frac{3}{2}}|\phi_p \rangle. \\  
\end{split}\ee
Considering the case where $c_M=0$, some coefficients would vanish after setting $h_M=0$ : $d_1=d_2=d_5=d_6=0$, and the null state is 
\be{} \begin{split}
|3/2\rangle&=d_3M_{-1}Q_{-\frac{1}{2}}|\phi_p \rangle+d_4M_{-1}\bar{Q}_{-\frac{1}{2}}|\phi_p \rangle. \\
\end{split}\ee

\paragraph{Level 2:} 
The general state is given by 
\be{} \begin{split}
&\Big[f_1L_{-2}+f_2L^2_{-1}+f_3L_{-1}M_{-1}+f_4M^2_{-1}+f_5M_{-2}+f_6L_{-1}Q_{-\frac{1}{2}}\bar{Q}_{-\frac{1}{2}}\\
&\hspace{1.7cm}+f_7M_{-1}Q_{-\frac{1}{2}}\bar{Q}_{-\frac{1}{2}}+f_8Q_{-\frac{3}{2}}\bar{Q}_{-\frac{1}{2}}+f_9Q_{-\frac{1}{2}}\bar{Q}_{-\frac{3}{2}}\Big]|\phi_p \rangle. \\  
\end{split}\ee
By similar techniques as done for the above cases ($c_M=0$) we can write down the null state at this level for 2 cases $h_L\neq1$ and $h_L=1$, after some coefficients drop out:
\be{} \begin{split}
|2\rangle&=\Big[f_4M^2_{-1}+f_7M_{-1}Q_{-\frac{1}{2}}\bar{Q}_{-\frac{1}{2}}\Big]|\phi_p \rangle \ \ \ (h_L\neq1)\\ 
&=\Big[f_6\Big(L_{-1}Q_{-\frac{1}{2}}\bar{Q}_{-\frac{1}{2}}-\frac{1}{2}Q_{-\frac{3}{2}}\bar{Q}_{-\frac{1}{2}}-\frac{1}{2}Q_{-\frac{1}{2}}\bar{Q}_{-\frac{3}{2}}\Big) \\
&\hspace{1.5cm}+f_4M^2_{-1}+f_7M_{-1}Q_{-\frac{1}{2}}\bar{Q}_{-\frac{1}{2}}\Big]|\phi_p \rangle \ \ \ (h_L=1).
\end{split}\ee
An analysis of the null states will reveal that $|1\rangle$ is just the descendant of the null state $|1/2\rangle$. Similarly $|3/2\rangle$ is the descendant of $|1\rangle$.  It is interesting to note that while descending from $|m\rangle$ to $|m+1/2\rangle$, we operate only with $Q^a_{-\frac{1}{2}}$ s. Also, to get from null state $|m\rangle$ to $|m+1\rangle$ we act twice with the $Q^a_{-\frac{1}{2}}$s as $$\{Q^a_{-\frac{1}{2}},Q^b_{-\frac{1}{2}}\}=\delta^{ab}M_{-1}.$$ We note that $Q^a_{-\frac{3}{2}}$ and $M_{-2}$ doesn't appear in $|3/2\rangle$ or $|2\>$ respectively. 

As the $Q$'s rotate amongst themselves to produce $M$'s, it would have been expected that setting $Q^a_{-r}|\phi_p \rangle=0$ for each level would truncate the algebra to a single copy of Virasoro. However in our analysis, we see that (up to level 2) we can neither set $Q_{-\frac{3}{2}}|\phi_p \rangle$ or $M_{-2}|\phi_p \rangle$ to zero. At the end of the day we must conclude that unlike $M_n$s in the case of the bosonic GCA \cite{Bagchi:2009pe}, or $M_n$s and $H_r$s in the case of SGCA$_I$, we cannot eradicate all of the $Q$'s in the SGCA$_H$. Thus the truncation from the SGCA$_H$ to a single copy of the Virasoro algebra is not possible.

\end{document}